\documentclass[
    ,final            % use final for the camera ready runs
  ]
  {aipproc}
 
\layoutstyle{6x9}

\newcommand{\newc}{\newcommand}

\newc{\MHp}{M_{H^\pm}}
\def\cphi{\phi}
\def\phiMe{\cphi_{M_1}}

\newc{\amuexp}{a_\mu^{\rm exp}}
\newcommand{\amutheo}{a_\mu^{\rm theo}}
\newcommand{\amu}{a_\mu}

\def\order#1{${\cal O}(#1)$}
\def\al{\alpha}
\newcommand{\mev}{\,\, \mathrm{MeV}}
\newcommand{\SZb}{\sin 2\beta\hspace{1mm}}
\newcommand{\als}{\alpha_s}
\def\de{\delta}
\def\De{\Delta}
\def\si{\sigma}
\newcommand{\gev}{\,\, \mathrm{GeV}}
\newcommand{\tb}{\tan \beta}
\newcommand{\mgl}{m_{\tilde{g}}}
\newcommand{\MA}{M_A}
\newcommand{\msusy}{m_{\tilde{f}}}%{M_{\SU}}
\newc{\Ga}{\Gamma}
\newc{\ie}{\ensuremath{\mathrm{i}}}
\newc{\eu}{\ensuremath{\mathrm{e}}}
\newc{\cw}[1][{}]{\ensuremath{c^{#1}_\mathrm{w}}}
\newc{\sw}[1][{}]{\ensuremath{s^{#1}_\mathrm{w}}}
\newc{\tw}[1][{}]{\ensuremath{\tan^{#1} \theta_{\mathrm{w}}}}
\newc{\ctw}[1][{}]{\ensuremath{\cot^{#1}\theta_{\mathrm{w}}}}
\newc{\real}{\ensuremath{\mathfrak{Re}}}
\newc{\imag}{\ensuremath{\mathfrak{Im}}}
\newc{\cv}[1][{}]{\ensuremath{\cos^{#1} \beta}}
\newc{\sv}[1][{}]{\ensuremath{\sin^{#1} \beta}}
\newc{\tv}[1][{}]{\ensuremath{\tan^{#1} \beta}}
\newc{\M}{\ensuremath{\mathcal{M}}}
\newc{\ssl}[1]{\ensuremath{\slashed{#1}}}
\newc{\half}{\ensuremath{\frac{1}{2}}}
\newc{\photino}{\widetilde{\gamma}}
\newc{\bino}{\widetilde{\cal B}}
\newc{\wino}{\widetilde{\cal W}}
\newc{\gluino}{\widetilde{\cal G}}
\newc{\rpv}{{\mbox{${\not\!\!R_p}$}}}
\newc{\cL}{{\cal L}}
\newc{\x}[1]{\ensuremath{\tilde{\chi}^0_{#1}}}
\newc{\GeV}{\ensuremath{\,\mathrm{GeV}}}
\newc{\fb}{\ensuremath{\,\mathrm{fb}}}
\newc{\pb}{\ensuremath{\,\mathrm{pb}}}
\newc{\erg}{\ensuremath{\,\mathrm{erg}}}
\def\lsim{\raise0.3ex\hbox{$\;<$\kern-0.75em\raise-1.1ex\hbox{$\sim\;$}}}
\def\gsim{\raise0.3ex\hbox{$\;>$\kern-0.75em\raise-1.1ex\hbox{$\sim\;$}}}
\newc{\MeV}{\,{\mathrm{MeV}}}
\newc{\lsp}{{{\tilde\chi}}}
\newc{\lam}{\lambda}
\newc{\ra}{\rightarrow}
\newc{\htext}[1]{{\color{red}  #1}}
\newc{\htextb}[1]{{\color{blue}  #1}}
\newc{\sweff}{\sin^2\theta_{\mathrm{eff}}}
\newc{\cha}[1]{\tilde \chi^\pm_{#1}}
\newc{\mcha}[1]{m_{\tilde \chi^\pm_{#1}}}
\newc{\neu}[1]{\tilde \chi^0_{#1}}
\newc{\mneu}[1]{m_{\tilde \chi^0_{#1}}}
\def\refeq#1{\mbox{Eq.~(\ref{#1})}}

\def\reffi#1{\mbox{Fig.~\ref{#1}}}

\def\refta#1{\mbox{Table~\ref{#1}}}

\def\citere#1{\mbox{Ref.~\cite{#1}}}
\def\citeres#1{\mbox{Refs.~\cite{#1}}}

\newcommand{\MZ}{M_Z}
\newcommand{\MW}{M_W}

\begin{document}

\title{Comments on a Massless Neutralino}


\author{H.~K.~Dreiner}{
  address={Bethe Center for Theoretical Physics \& Physics Institute, 
University of Bonn, Bonn, Germany}
}

\begin{abstract}
  We consider first an interesting connection between the development
  of physics and the Boston Red Sox. We then discuss in detail the
  collider phenomenology, as well as precision electroweak observables
  of a very light neutralino. We conclude by considering also the
  astrophysics and cosmology of a very light neutralino. We find that
  a massless neutralino is consistent with all present data.
\end{abstract}

\maketitle

%%%%%%%%%%%%%%%%%%%%%%%%%%%%%%%%%%%%%%%%%%%%
%% MAINMATTER
%%%%%%%%%%%%%%%%%%%%%%%%%%%%%%%%%%%%%%%%%%%%

\section{Preface}

\begin{figure}[t]
\begin{picture}(200,510)
\put(-155,187){\includegraphics{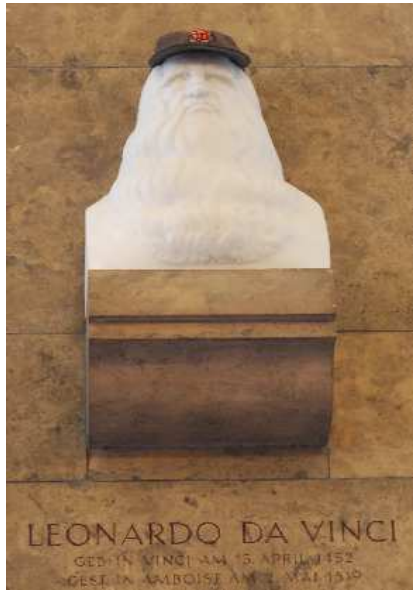}}
\put(50,180){\includegraphics{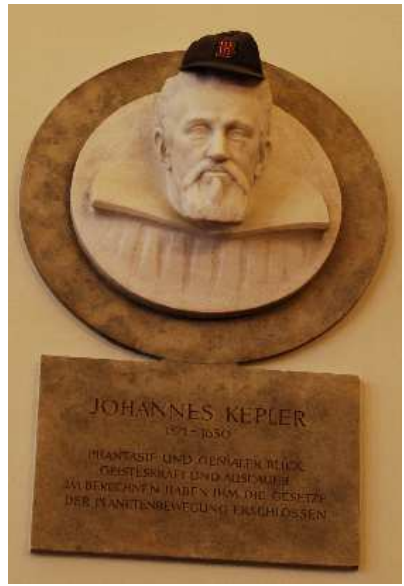}}
\put(-170,-125){\includegraphics{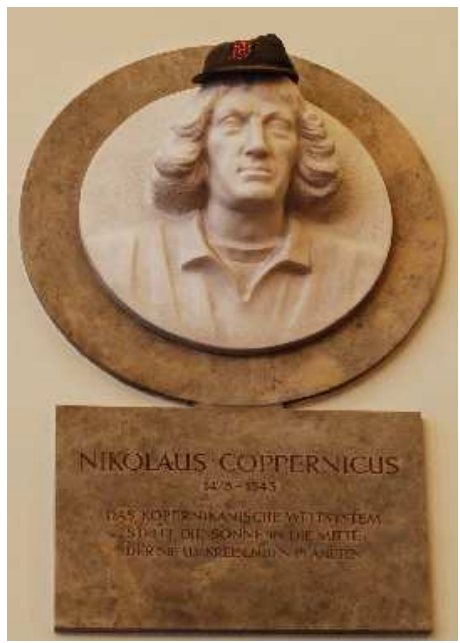}}
\put(10,-135){\includegraphics{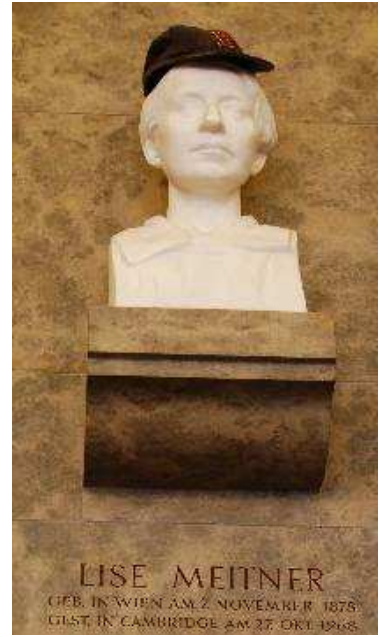}}
\put(-20,255){(a)}
\put(210,255){(c)}
\put(-20,-8){(b)}
\put(210,-8){(d)}
\end{picture}
\caption{\small (a) Leonardo Davinci, (b) Nikolaus Copernicus (c) Johannes Kepler, 
and (d) Lise Meitner showing their support.}
\label{davinci-kepler}
\end{figure}

\begin{figure}[t!]
\begin{picture}(200,270)
\put(-200,-135){\includegraphics{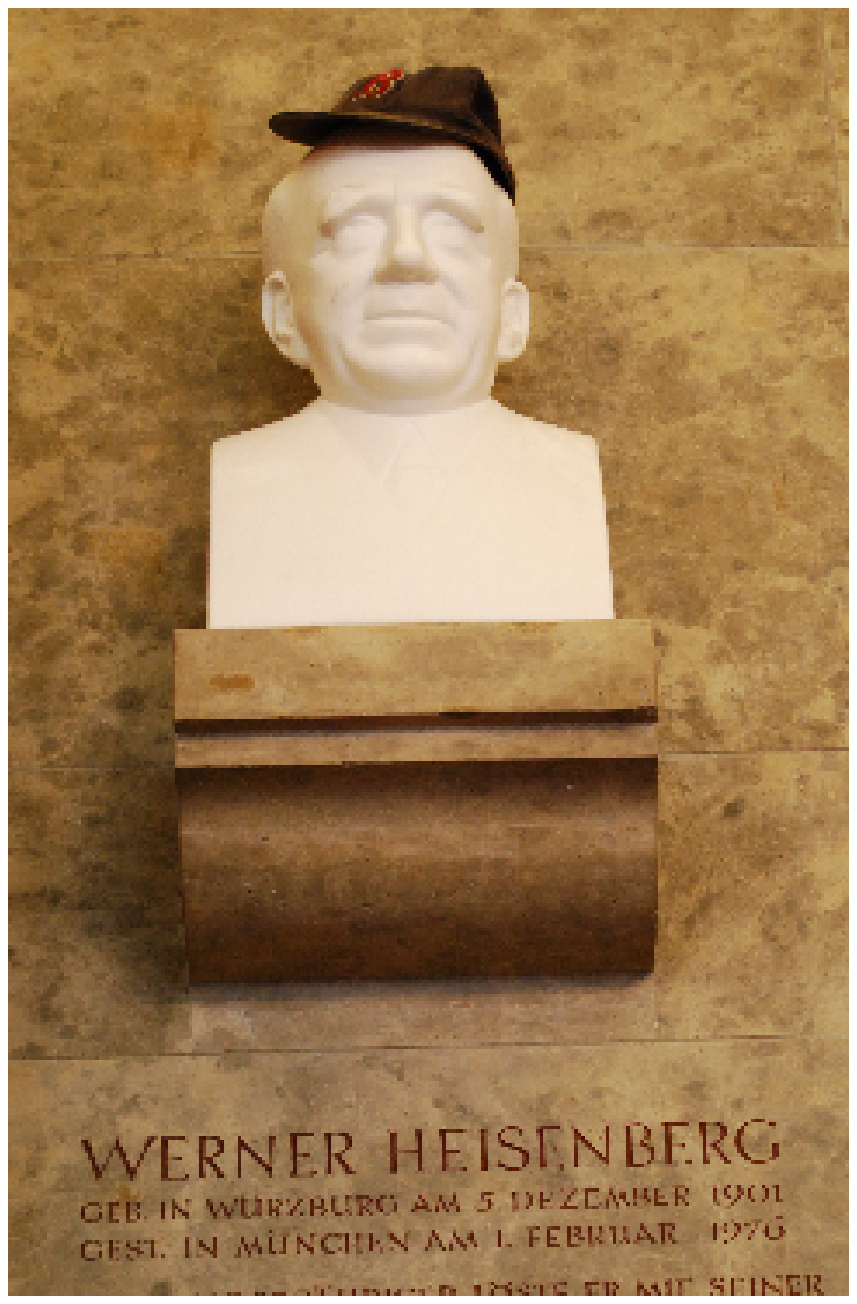}}
\put(0,-160){\includegraphics{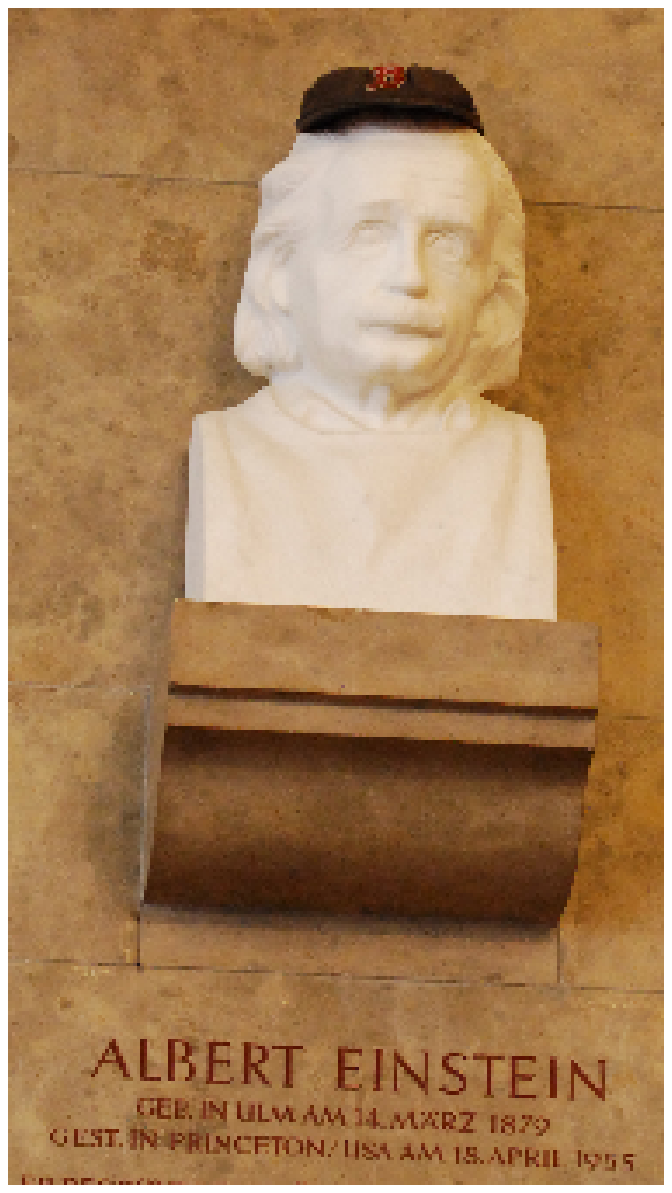}}
\put(78,3){(a)}
\put(108,3){(b)}
\end{picture}
\caption{\small Also (a) Werner Heisenberg and (b) Albert Einstein are dedicated followers. }
\label{heisenberg-einstein}
\end{figure}

It is a pleasure to speak here at Northeastern University, in Boston.
I was born not too far, in Williamstown, some time ago, but have lived
in Europe for quite some time. Now, it is great to be back in
Massachusetts.  There is of course a longstanding connection between
scientific circles in Europe and this wonderful city and state. It is
even rumored that Galileo Galilei (1564--1642) was offered a job next
door at Harvard (est. 1635).  However, Galileo Galilei never came,
possibly because he wasn't sure of getting tenure, or possibly because
of earlier support for Boston from fellow Italian Leonardo Davinci
(1452--1519), see Fig.~\ref{davinci-kepler}a. Galieli always did want
to be different.  This strong affiliation to Boston has been
maintained through the years by Nicolaus Copernicus (1473--1543),
Fig.~\ref{davinci-kepler}b, and Johannes Kepler (1571--1630) who
worked in T\"ubingen and Prag but dreamt of Red Sox nation,
Fig.~\ref{davinci-kepler}c. In more modern times Lise Meitner
Fig.~\ref{davinci-kepler}d and Werner Heisenberg,
Fig.~\ref{heisenberg-einstein}a, have upheld the torch. In the
20$^{\mathrm{th}}$ century, the culmination of this connection in the
person of Albert Einstein, Fig.~\ref{heisenberg-einstein}b, is well
documented.

\vspace{1cm}

%%%%%%%%%%%%%%%%%%%%%%%%%%%%%%%%%%%%%%%%%%%%%%%%%%%%%%%%%%%%%%%%%%%%%%%%%%
%%%%%%%%%%%%%%%%%%%%%%%%%%%%%%%%%%%%%%%%%%%%%%%%%%%%%%%%%%%%%%%%%%%%%%%%

\hspace{2cm}\begin{minipage}{12cm}
\begin{enumerate}

\item[1903\phantom{+}] Red Sox win 1$^{\mathrm{st}}$ ever World Series 
on grounds of Northeastern

\item[1904\phantom{+}] Red Sox win AL pennant3
\item[1905\phantom{+}] ---- Einstein, inspired, has the season of his life 

\item['10's\phantom{+}] Red Sox dominant force

\item[1916\phantom{+}] ---- Einstein: General Relativity

\item[1918\phantom{+}] Sox win last World Series for a while

\end{enumerate}
\end{minipage}

\hspace{2cm}\begin{minipage}{12cm}
\begin{enumerate}
\item[1919\phantom{+}] Babe Ruth Sold

\item[1919\phantom{+}] ---- Einstein, past his prime, slumps

\item[2004\phantom{+}] Red Sox win Series again, hurrah!

\item[2005\phantom{+}] World celebrates Einstein centennial, double hurrah!

\item[2005+] -- Both Red Sox and Einstein in 7$^{\mathrm{th}}$ heaven

\end{enumerate}
\end{minipage}

\section{Introduction}

The lightest supersymmetric particle, the LSP, plays a special role in
the search for supersymmetry at colliders. It is the end product of
the cascade decay of any produced SUSY particle. Thus the nature of
the LSP is decisive for all supersymmetric signatures at the LHC. For
conserved proton hexality \cite{Dreiner:2005rd,Dreiner:2007vp} the LSP
is stable and must be the lightest neutralino: $\neu{1}$. Here enquire:
`How light can the $\neu{1}$ be?'. This is a summary of previous work
\cite{Choudhury:1999tn,Dreiner:2003wh,Dreiner:2006sb,Dreiner:2007vm,Dreiner:2007fw,Dreiner:2009ic}.
%%%%%%%%%%%%%%%%%%%%%%%%%%%%%%%%%%%%%%%%%%%%%%%%%%%%%%%%%%%%%%%%%%%%%%%%%%
%%%%%%%%%%%%%%%%%%%%%%%%%%%%%%%%%%%%%%%%%%%%%%%%%%%%%%%%%%%%%%%%%%%%%%%%%%
The PDG cites as the laboratory bound
\cite{Amsler:2008zz}
\begin{equation}
\mneu{1}>46\GeV    %,\quad @\, 95\%\;\mathrm{C.L.}\,,
\label{lep-bound}
\end{equation}
at 95\% C.L., which is based on the chargino searches at LEP. These
yield lower limits on $M_2$ and $\mu$.  Furthermore, this bound
assumes an underlying SUSY GUT, \textit{i.e.} $M_1 =
\frac{5}{3}\,\tw[2]\; M_2 \approx \half M_2\,.$ The experimental bound
on $M_2$ then implies a lower bound on $M_1$.  Using the neutralino
mass matrix, together these give rise to the lower bound in
Eq.~(\ref{lep-bound}).

It is the purpose of this paper to investigate the consequences of
dropping the SUSY GUT assumption. In this more general scenario, $M_1$
and $M_2$ are both free parameters. We systematically demonstrate that
then a \textit{massless} neutralino is consistent with theory and all
present laboratory data. For possible models see
\cite{Chamseddine:1995gb,Dudas:2008eq} and techniques
\cite{Dreiner:2008tw}. Taking the determinant of the neutralino mass
matrix \cite{Dreiner:2009ic} and setting it to zero, we get
\begin{equation}
\mu\,\big[\,M_2 \MZ^2\sw[2]\sin(2\beta) + M_1\big(-M_2 \mu 
+ \MZ^2\cw[2]\sin(2\beta)\big)\,\big]=0\,.
\end{equation}
The solution $\mu=0$ is excluded by the LEP chargino bounds. Solving
for $M_1$ yields
\begin{eqnarray}
M_1 = \frac{M_2 \MZ^2 \sin(2\beta)\sw[2]}{\mu M_2-\MZ^2\sin(2\beta)\cw[2]}\,.
\label{massless-neut-condition}
\end{eqnarray}
Thus for every value of $M_2,\,\mu$ and $\tan\beta$, we can find a
value of $M_1$ with $M_{\tilde\chi^0_1}=0$. This is stable under radiative
corrections \cite{Dreiner:2009ic}. In the case of complex gaugino
parameters there is not always a solution. In the real case,
Eq.~(\ref{massless-neut-condition}) leads to $M_1\approx M_2/40$. We
find that for very light neutralinos they are typically more than 90\%
bino.

\section{Collider Bounds}

\begin{figure}[t]
\begin{picture}(200,180)
\put(-110,-0){\includegraphics{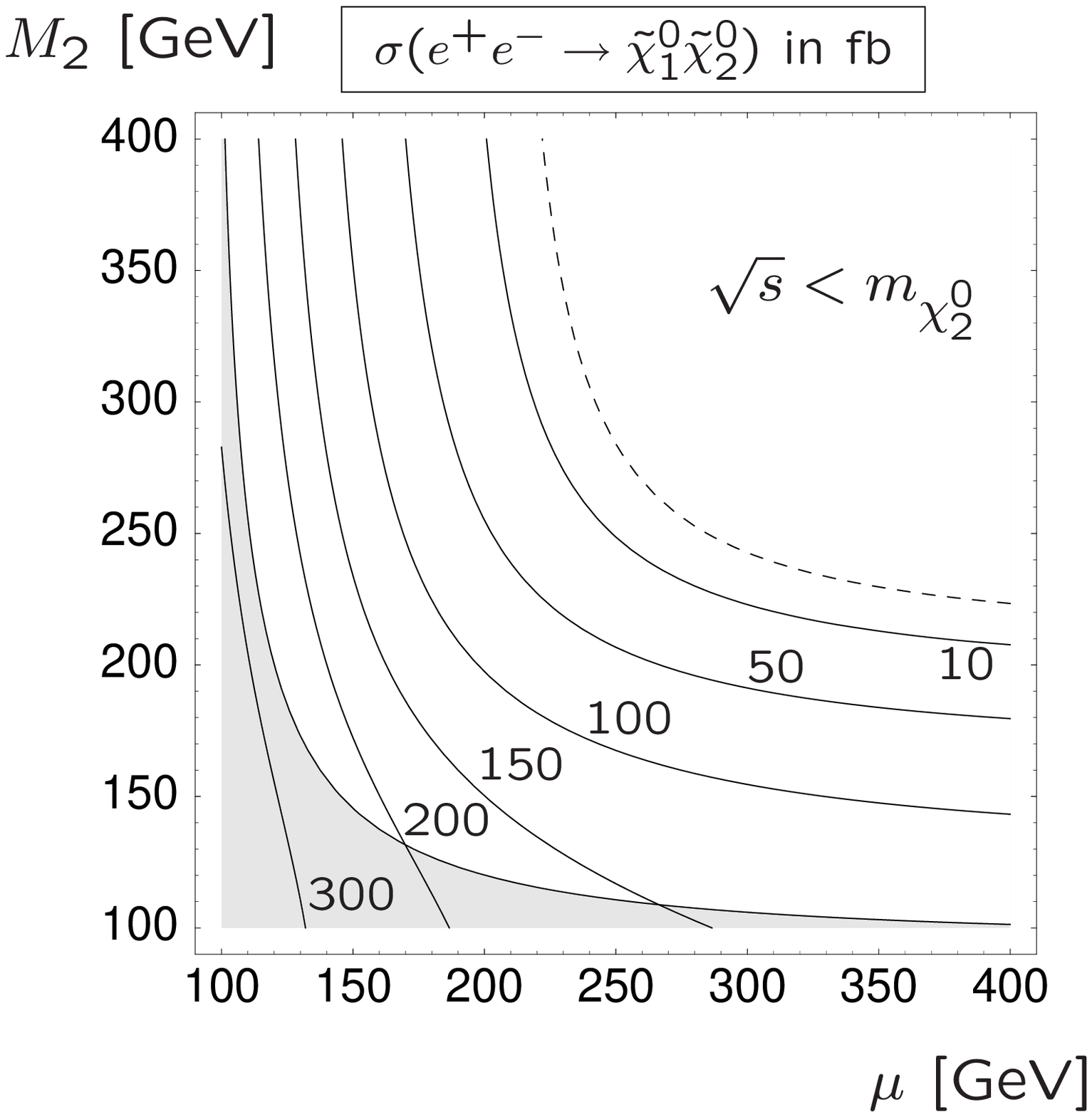}}
\put(100,-30){\includegraphics{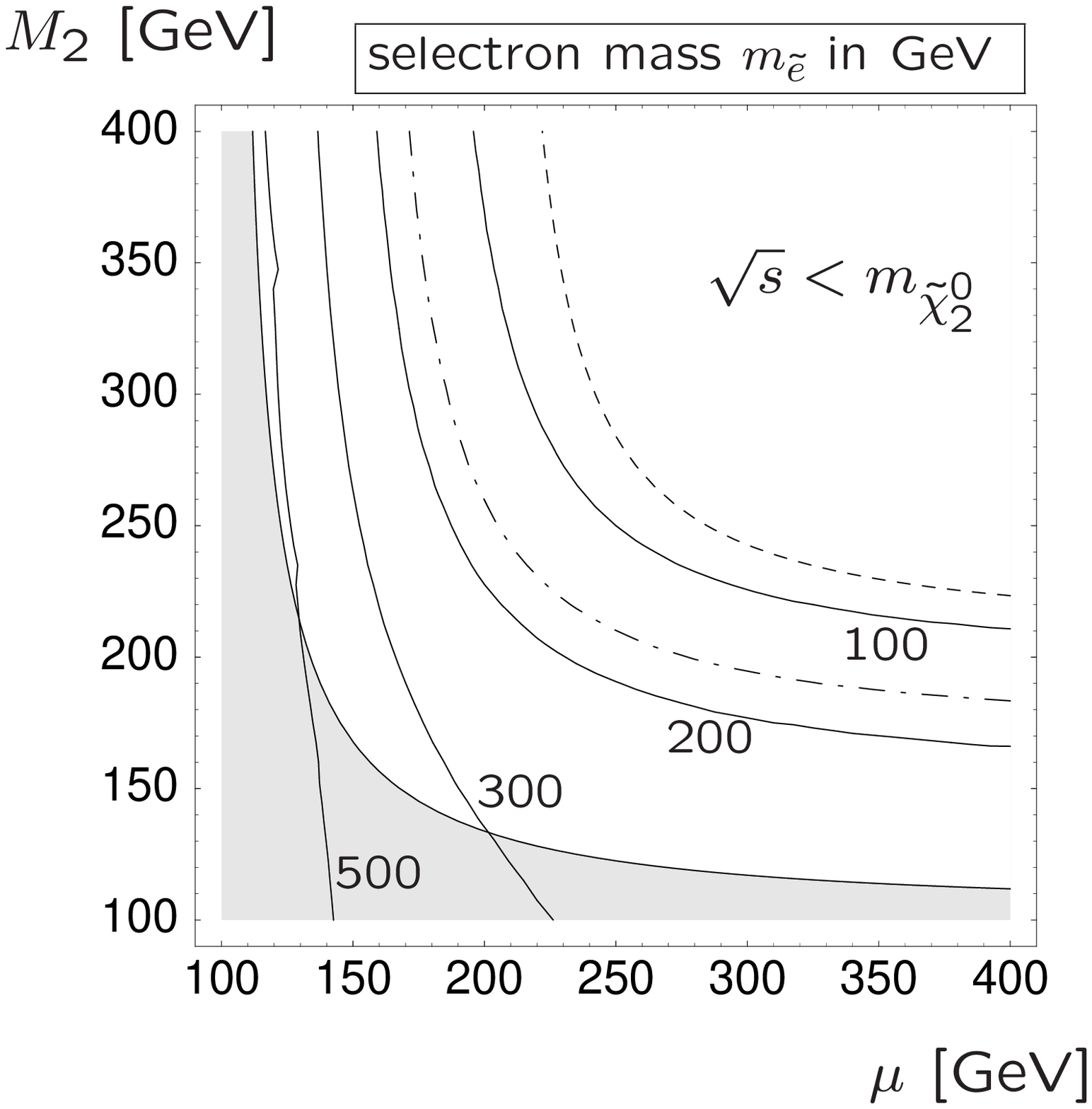}}
\put(0,0){\small\bf (a)}
\put(230,0){\small\bf (b)}
\end{picture}
\caption{\small {\bf (a)} Contour lines of $\sigma(e^+e^-\to\neu{1}
  \neu{2})$ with $\tan\beta=10$, and $m_{\tilde e}=200$~GeV, at $\sqrt
  s = 208$~GeV.  {\bf (b)} Lower bounds on the selectron mass
  $m_{\tilde e}$, such that
  $\sigma(e^+e^-\to\neu{1}\neu{2})=70\,\mathrm{fb}$. }
\label{fig:OPALbounds}
\end{figure}

\textbf{Neutralino production at LEP:} If we assume $\mneu{1} =0$ the
associated production, $e^+e^-\to\neu{1}\neu{2}$, would be accessible
at LEP up to the kinematical limit of $\sqrt s=\mneu{2} = 208 \GeV$.
In order to compare with the results of the LEP searches we make use
of the model-independent upper bounds on the topological neutralino
production cross section obtained by OPAL with $\sqrt s=
208$~GeV~\cite{Abbiendi:2003sc},
\begin{equation}
\sigma(e^+e^-\to\neu{1}\neu{2})\times
{\rm BR}(\neu{2}\to Z\neu{1})\times {\rm BR}(Z\to q\bar q) .
\end{equation}
Taking into account ${\rm BR}(Z\to q\bar q)\approx70\%$, one can
read off from the OPAL plots \cite{Abbiendi:2003sc},
\begin{equation}
\sigma(e^+e^-\to\neu{1}\neu{2}) \times {\rm BR}(\neu{2}\to Z\neu{1})
<70\,\mathrm{fb}\,.  \label{bound}
\end{equation}
We analyze this bound assuming conservatively that ${\rm BR}(\neu{2}
\to Z\neu{1}) = 1$. Imposing the bound, Eq.~(\ref{bound}),
significantly constrains the parameter space. In
Fig.~\ref{fig:OPALbounds}(a) we show contour lines of the cross
section $\sigma(e^+e^- \to \neu{1}\neu{2})$ in the $\mu$--$M_2$ plane
for $\tan\beta = 10$ and degenerate selectron masses $m_{\tilde
  e_R}=m_{\tilde e_L}=m_{\tilde e}=200$~GeV.  We observe that there is
a large region where $\sigma>70$~fb. Here the selectron masses have to
be sufficiently heavy. Thus, the bound on the neutralino production
cross section can be translated into lower bounds on the selectron
mass $m_{\tilde e}$, for $\mneu{1}=0$.  In
Fig.~\ref{fig:OPALbounds}(b), we show contours of the selectron mass,
such that the bound $\sigma(e^+e^-\to\neu{1}\neu{2}) < 70$~fb is
fulfilled.

%%%%%%%%%%%%%%%%%%%%%%%%%%%%%%%%%%%%%%%%%%%%%%%%%%%%%%%%%%%%%%%%%%%%%%%%%%%%%%
%%%%%%%%%%%%%%%%%%%%%%%%%%%%%%%%%%%%%%%%%%%%%%%%%%%%%%%%%%%%%%%%%%%%%%%%%%%%%

\textbf{Radiative neutralino production:} An additional search channel
is radiative neutralino production, $e^+e^-\to\neu{1}\neu{1}\gamma$.
Due to the large SM background, $e^+e^-\to\nu\bar\nu\gamma$, the
significance is at best $S\approx0.1$ for $\mathcal L=100$~pb$^{-1}$
and $\sqrt s=208\,$GeV~\cite{Dreiner:2006sb,Dreiner:2007vm}. Cuts on
the photon energy or angle do not help, due to similar distributions
of signal and background.  We find a similar situation at
$b$-factories, $\sqrt s \approx 10$~GeV. An identification of the
signal `photon plus missing energy' is difficult due to the large
photonic background from the abundant hadronic processes.  At the ILC,
radiative neutralino production would be measurable, due to the option
of polarised
beams~\cite{Choudhury:1999tn,Dreiner:2006sb,Dreiner:2007vm,MoortgatPick:2005cw}.

\section{Precision Observables}

\begin{figure}
\begin{picture}(200,250)
\put(-90,-0){\includegraphics{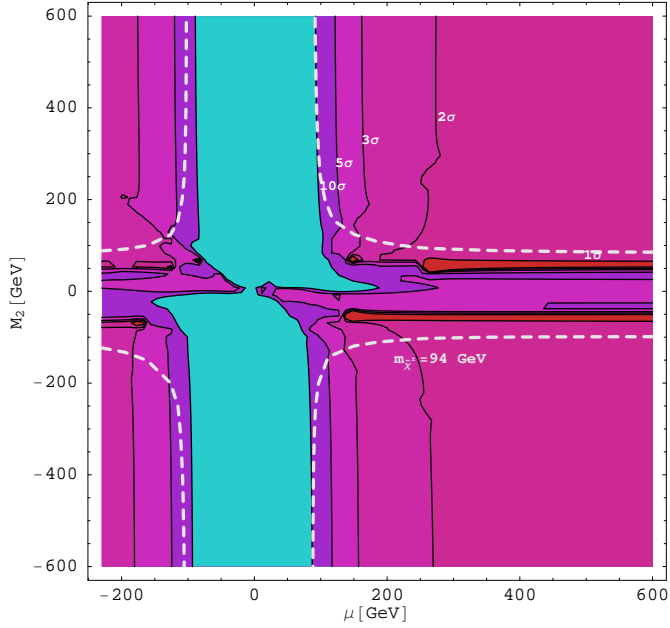}}
\end{picture}
\caption{The difference of the experimental value and the theory 
  prediction for the invisible $Z$~width, $\de\Ga_{\rm inv}$,
  indicated as $\de\Ga_{\textup{inv}}\equiv(\Ga_{\textup{inv}}-\Ga^{
  \textup{exp}}_{\textup{inv}}) = (10,5,3,2,1)\times\sigma^{\textup{
  exp}}_{\Ga_{\textup{inv}}}$ contours.}
\label{fig:MUEM2plane} 
\vspace{-2em}
\end{figure}

In the following we study the impact of a light or massless neutralino
on electroweak precision physics. We consider the full one-loop and
leading higher-order corrections
\cite{dr2lA,drMSSMal2A,drMSSMal2B,Heinemeyer:2006px}. We focus on the
invisible $Z^0$ width, $\Gamma_{\textup{inv}}$, as an example. The
additional contributions due to
$\Ga_{\neu{1}}=\Gamma(Z\to\neu{1}\neu{1})$ can be large if the
neutralino has a considerable Higgsino component. In
\citere{ZObsMSSM}, the processes~$Z\to\neu{1}\neu{1}$ and $Z\to f \bar
f$ have been calculated at~$\mathcal{O}({\alpha})$ and supplemented
with leading higher-order terms from the SM and the MSSM. The
experimental values for the total width and the invisible width of the
$Z$ boson are~\cite{Amsler:2008zz}
\begin{equation}
\Ga_Z^{\rm exp} = 2495.2 \pm 2.3 \,\mathrm{MeV}, \qquad 
\label{GaZinv}
\Ga_{\rm inv}^{\rm exp} =499.0 \pm 1.5 \,\mathrm{MeV}~.  
\end{equation}
%Below, we label the experimental errors of these two quantities as
%$\si_{\Ga_Z}^ {\rm exp}$ and $\si_ {\Ga_{\rm inv}}^{\rm exp}$, respectively. 
In our numerical analysis, we show the results for
\begin{eqnarray}
\label{deGaZinv}
\de\Ga_{\rm inv} &\equiv& \Ga_{\rm inv} - \Ga_{\rm inv}^{\rm exp}\,,
% \\
%\label{deGaZ}
%\de\Ga_Z &\equiv& \Ga_Z - \Ga_Z^{\rm exp} ,
\end{eqnarray}
%\textit{i.e.} the difference of the MSSM
%prediction and the experimental result for the invisible $Z$~width,
%$\de\Ga_{\textup{inv}}$.
We investigate $\de\Ga_{\rm inv}$ in one representative SUSY parameter
regions. We choose fairly light scalar fermions and set the diagonal
soft SUSY-breaking parameter $\msusy$ to $250\gev$. In
\reffi{fig:MUEM2plane}, we show $\de\Ga_{\rm inv}$ as a function of
$M_2$ and $\mu$. $M_1$ is fixed via
Eq.~(\ref{massless-neut-condition}).  The remaining SUSY parameters
are $\tb=10$, $A_{\tau}=A_t=A_b=\mgl=\MA= 500\gev$.  Here $A_f$
denotes the trilinear couplings of the Higgses to the sfermions,
$\mgl$ is the gluino mass, and $\MA$ the mass of the CP-odd Higgs
boson. The deviations from the experimental central values as given in
Eqs.~(\ref{deGaZinv}), are indicated as experimental
$n\times\sigma$~contours of the respective observable.
%%%%%%%%%%%%%%%%%%%%%%%%%%%%%%%% Begin FIGURE %%%%%%%%%%%%%%%%%%%%%%%%%%%%%%%%%
%%%%%%%%%%%%%%%%%%%%%%%%%%%%%%%% End FIGURE %%%%%%%%%%%%%%%%%%%%%%%%%%%%%%%%%%%
In addition, the $95\%\,$C.L.\ exclusion bounds of $\mcha{1}>94\gev$
\cite{Amsler:2008zz} on the chargino mass from direct searches are
marked by dashed white lines. Fig.~\ref{fig:MUEM2plane} clearly
displays that for both observables the MSSM prediction can deviate
considerably from the experimental values. This is in particular the
case for small $|\mu|$ and small $|M_2|$. Nearly all of the parameter
space ruled out at the $5\sigma$~level for $\Ga_{\textup{inv}}$ is,
however, already excluded due to direct chargino searches. For the
interpretation of these plots it is furthermore important to keep in
mind that the results for $\Ga_{\textup{inv}}$ do not only depend on
$\mu$ and $M_2$, but on all the other SUSY parameters as well. This
means in particular that an apparent $1 \sigma$ effect can easily be
caused or canceled out by, for instance, a change induced by $\msusy$,
which is known to have a strong impact on the decay into SM fermions,
see also the discussion in Ref.~\cite{ZObsMSSM}.  Furthermore even in
the SM, $\Ga_{\textup{inv}}$ is predicted to be slightly larger than
the experimentally measured value, resulting in a $\sim 1 \si$
deviation. In summary, $\Ga_{\rm inv}$ cannot exclude a massless
neutralino. The parts of the $\mu$-$M_2$~planes that lead to a large
deviation from the experimental values are mostly already excluded by
direct chargino searches.

\section{Electric Dipole Moments and Rare Meson Decays}

For the electric dipole moments we refer the reader to the original
paper \cite{Dreiner:2009ic}.  For the rare meson decays we defer to
the dedicated talk by Ben O'Leary (RWTH Aachen) also given at this
conference \cite{O'Leary:2009zc,Dreiner:2009er}.  For some relevant
techniques see \cite{Dreiner:2001kc,Dreiner:2006gu}

\section{Astrophysics and Cosmology}

\textbf{Supernova Cooling:} A very light neutralino, $M_\chi\leq
\mathcal{O}(100\,\mathrm{MeV})$
can contribute to supernova cooling
\cite{Dreiner:2003wh,Grifols:1988fw,Ellis:1988aa}. The
two main production mechanisms are
\begin{equation}
e^+ + e^-\longrightarrow  \neu{1} + \neu{1}\,, \qquad\;
N+N\longrightarrow N+N+\neu{1}+\neu{1}\,. \label{ul:Xs-n}
\end{equation}
Once produced, the neutralinos have a mean-free-path, $\lam_{\neu{1}}$,
in the supernova core which is determined via the cross sections for 
the processes
\begin{equation}
\neu{1} +e \longrightarrow \neu{1} + e\,,\qquad\quad
\neu{1} + N \longrightarrow \neu{1} + N\,,
\label{ul:scatter2}
\end{equation}
\begin{figure}[t!]
\begin{picture}(200,240)
\put(-110,240){\includegraphics{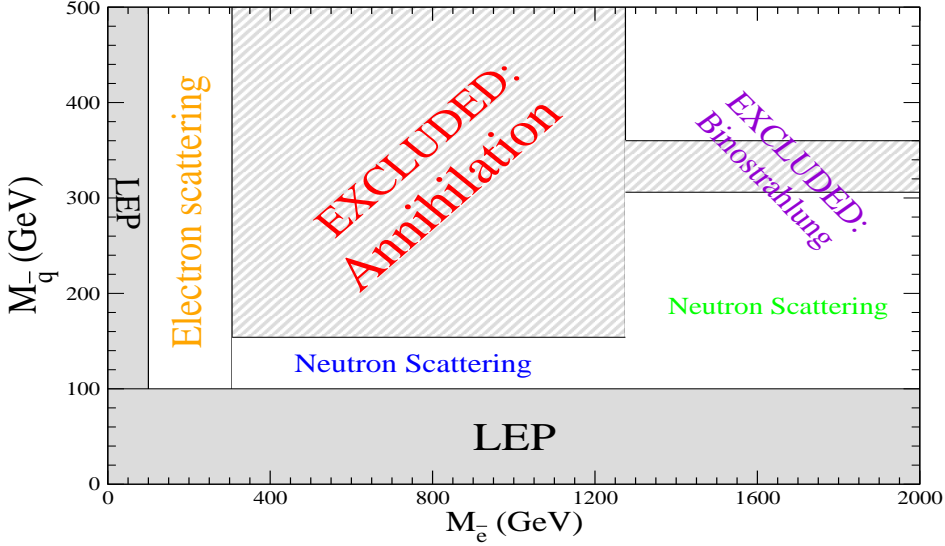}}
\end{picture}
\caption{\small Exclusion in SUSY plane for a massless neutralino from supernova cooling \cite{Dreiner:2003wh}. }
\label{supernova}
\end{figure}
In order to retain the observed neutrino signal we obtain the
exclusion regions depicted in Fig.~\ref{supernova}. These depend
strongly on the selectron and squark masses which enter in the
propagators of the processes Eqs.~(\ref{ul:Xs-n}), (\ref{ul:scatter2}).

\textbf{Hot Dark Matter, the Cowsik--McClelland Bound:} 
Here, we consider the case of a (nearly) massless neutralino,
$\mneu{1}\lsim\mathcal{O}(1\, \mathrm{eV})$. Since the very light bino
contributes to the hot dark matter of the universe, we assume here
implicitly that the cold dark matter originates from another
source. The bino relic energy density, $\rho_{\bino}$,
divided by the critical energy density of the universe, $\rho_c$, is
given by \cite{Kolb:1990eu}
\begin{eqnarray}
\label{ul:eq:relicdensity}
\Omega_{\bino} &\equiv& \frac{\rho_{\bino}}{\rho_c}
\;=\;  \frac{43}{11}\,\zeta(3)\,\frac{8\pi G_N}
{3 H_0^2}\,\frac{g_{\mathrm{eff}}(\bino)}{g_{\ast S}(T)}\,
            T_\gamma^3 \,m_{\bino}\,.
\end{eqnarray}
In order for the bino hot dark matter not to disturb the structure
formation, we assume its contribution to be less than the upper bound
on the energy density of the neutrinos, as determined by the WMAP data
\cite{Dunkley:2008ie}
\begin{equation}
\label{eq:bound}
\Omega_{\bino} h^2 \le  [\Omega_{\nu} h^2]_{\mathrm{max}} = 0.0076\,.
\end{equation}
{}From Eqs.~(\ref{ul:eq:relicdensity}) and (\ref{eq:bound}), we 
find the conservative upper bound
\begin{eqnarray}
m_{\bino}  \le 0.07 \enspace\mathrm{eV}\,.
\end{eqnarray}
Thus a very light bino with mass below about 0.1 GeV is consistent
with structure formation. This line of argument was originally used by
Gershtein and Zel'dovich \cite{Gershtein:1966gg} and Cowsik and
McClelland~\cite{Cowsik:1972gh} to derive a neutrino upper mass bound,
by requiring $\Omega_\nu\leq1$. We have here obtained an upper
mass bound for a hot dark matter bino.

\textbf{Acknowledgments} I would like to thank Christoph Hanhart, 
Sven Heinemeyer, Olaf Kittel, Ulrich Langenfeld, Daniel Phillips and
Georg Weiglein for the collaborations underlying this work. This work was
partially supported by SFB TR-33 The Dark Universe. I thank the
Deutsche Museum, M\"unchen, for hospitality during a
\textbf{``Physikshow Bonn''} visit, in March 2009.

%\vspace{-1cm}

%%%%%%%%%%%%%%%%%%%%%%%%%%%%%%%%%%%%%%%%%%%%%%%%%%%%%%%%%%%%%%%%%%%%%%%%%%
\begin{figure}[h!]
\begin{picture}(200,202)
\put(-40,-80){\includegraphics{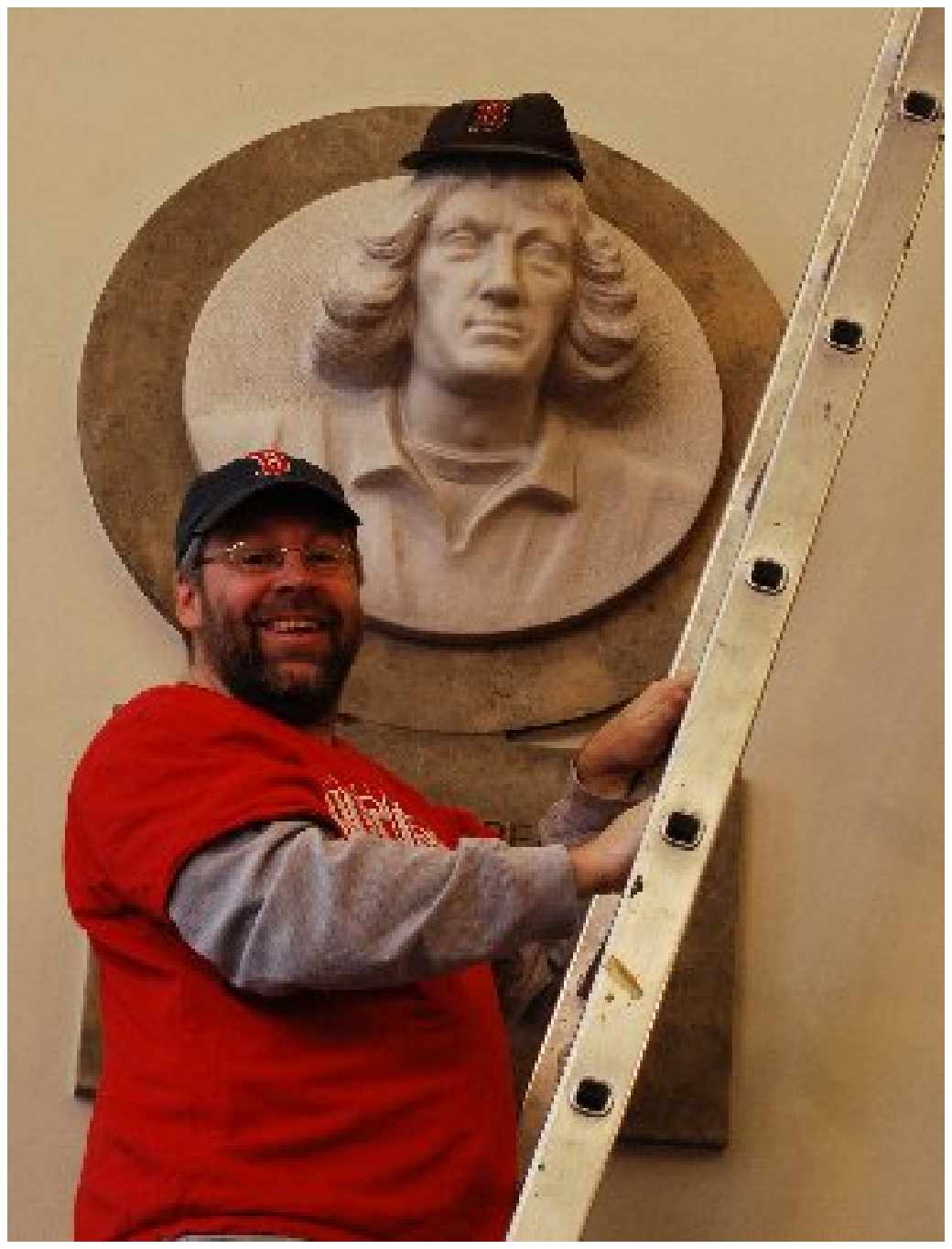}}
\end{picture}
%\caption{\small (a) Werner Heisenberg and (b) Albert Einstein. }
%\label{heisenberg-einstein}
\end{figure}
%%%%%%%%%%%%%%%%%%%%%%%%%%%%%%%%%%%%%%%%%%%%%%%%%%%%%%%%%%%%%%%%%%%%%%%%%%

\end{document}